\documentclass[11pt]{article}
\usepackage{amsmath}
\usepackage{graphicx}

\newcommand{\rme}{\mathrm{e}} %
\newcommand{\rmd}{\mathrm{d}} %
\newcommand{\eref}[1]{(\ref{#1})}%
\newcommand{\Eref}[1]{Equation~(\ref{#1})}%
\newcommand{\fref}[1]{figure~\ref{#1}} %
\newcommand{\Fref}[1]{Figure~\ref{#1}}%
\newcommand{\vect}[1]{\overrightarrow{#1}}%

\newcommand{\MDP}{\hbox{MDP--$p$\ }} %
\newcommand{\MDPns}{\hbox{MDP--$p$}}%
\newcommand{\dilog}{\mathrm{Li_2}}%

\begin{document}

\title{A note on limit shapes of minimal difference partitions}

\author{Alain Comtet$^{1,2}$, Satya N. Majumdar$^{1}$, and Sanjib
  Sabhapandit$^{1}$}

\date{$^{1}$Laboratoire de Physique Th\'eorique et Mod\`eles Statistiques,
  Universit\'e de Paris-Sud, CNRS UMR 8626, 91405 Orsay
  Cedex, France \\
  $^{2}$ Institut Henri Poincar\'e, 11 rue Pierre et Marie Curie, 75005
  Paris, France}
  
  \maketitle

\begin{abstract}
  We provide a variational derivation of the limit shape of minimal
  difference partitions and discuss the link with exclusion statistics.
\end{abstract}

{\it This paper is dedicated to Professor Leonid Pastur for his 70th
  anniversary}.

\bigskip

\begin{figure}
\centering
\includegraphics[height=7cm]{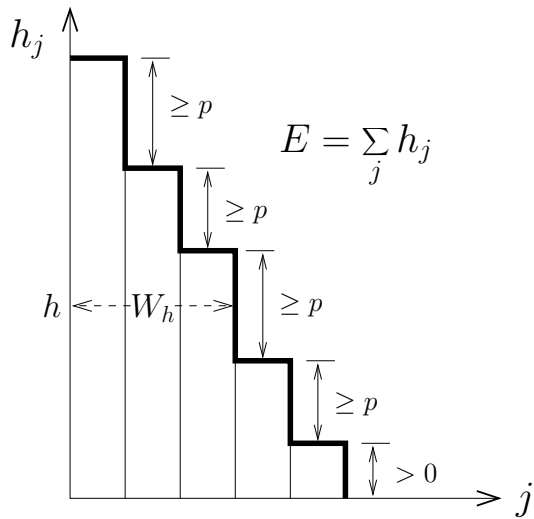}
\caption{\label{mdp} A typical Young diagram for \MDP problem. The thick
  solid border shows the height profile or the outer perimeter. $W_h$ is
  the width of the Young diagram at a height $h$, i.e., $W_h$ is the number
  of columns whose heights $\ge h$.}
\end{figure}

A partition of a natural integer $E$~\cite{Andrews} is a decomposition of
$E$ as a sum of a nonincreasing sequence of positive integers $\{h_j\}$,
i.e., $E=\sum_j h_j$ such that $h_j \ge h_{j+1}$, for $j=1,2\ldots$.  For
example, $4$ can be partitioned in $5$ ways: $4$, $3+1$, $2+2$, $2+1+1$,
and $1+1+1+1$.  Partitions can be graphically represented by Young diagrams
(also called Ferrers diagrams) where $h_j$ corresponds to the height of the
$j$-th column. The $\{h_j\}$'s are called the parts or the summands of the
partition. One can put several constraints on such partitions.  For
example, one can take the number of columns $N$ to be fixed or put
restrictions on the heights.  In this paper we focus on a particular
constrained partition problem called the minimal difference $p$ partitions
(\MDPns).  The \MDP problem is defined by restricting the height difference
between two neighboring columns, $h_{j} -h_{j+1} \ge p$.  For instance the
only allowed partitions of $4$ with $p=1$ are $4$ and $3+1$. A typical
Young diagram for \MDP problem is shown in \fref{mdp}.  Consider now the
set of all possible partitions of $E$ satisfying $E=\sum_j h_j$ and
$h_j-h_{j+1} \ge p$. Since this is a finite set, one can put a uniform
probability measure on it, which means that all partitions are
equiprobable. Then, a natural question is: what is the typical shape of a
Young diagram when $E\rightarrow\infty$?

In the physics literature this problem was first raised by Temperley, who
was interested in determining the equilibrium profile of a simple cubic
crystal grown from the corner of three walls at right angles.  The two
dimensional version of the problem ---where walls (two) are along the
horizontal and the vertical axes and $E$ ``bricks'' (molecules) are packed
into the first quadrant one by one such that each brick, when it is added,
makes two contact along faces--- corresponds to the $p=0$ partition
problem.  Temperley~\cite{Temperley} computed the equilibrium profile of
this two dimensional crystal.  In the mathematics literature the
investigation of the shape of random Young tableaux was started by Vershik
and Kerov \cite{Kerov} and independently by Logan and Shepp \cite{Logan}.
The case of uniform random partitions was treated by Vershik and
collaborators~\cite{Vershik, Vershik2, Vershik3} who obtained the limit
shapes for the $p=0$ and $p=1$ cases and also the average deviations from
the limit shapes~\cite {Dembo}.  Some of these results were extended by
Romik~\cite{Romik} to the \MDP for $p=2$. These problems belong to the
general framework of asymptotic combinatorics, a subject which displays
unexpected links with random matrix theory \cite{P}.  In this note we
compute the limit shapes of \MDP for all $p\ge 0$ by a variational approach
and mention an interesting link with exclusion statistics.

We first recapitulate the arguments used in~\cite{Shlosman, PLK} (see also
\cite{RD} for a similar approach) to compute the limit shapes of the Young
diagrams of unrestricted partitions ($p=0$).  Let $P=(i,h_i)$ and
$Q=(j,h_j)$ be two points belonging to the outer perimeter of the Young
diagram of a given partition.  We evaluate the total number of subdiagrams
which connects these two points.  These subdiagrams are lattice staircases
with the only restriction that each step either goes right or downward.
The total number of horizontal steps is $j-i$, the total number of vertical
steps is $h_i-h_j$, and the total number of steps is $j-i+h_i-h_j$.
Therefore, the total number of configurations is
\begin{equation}
  \Omega_0(P,Q)\equiv \Omega_0(i,h_i;j,h_j)=\binom{j-i+h_i-h_j}{j-i}.
\label{omega0}
\end{equation}
If $P$ and $Q$ are far apart (i.e., $a=j-i \gg 1, b=h_i-h_j\gg 1$) we may
use the Stirling formula which gives
\begin{equation}
  \ln \Omega_0 (P,Q) = -a\ln \frac{a}{a+b} -b\ln \frac{b}{a+b}
  =\sqrt{a^2+b^2}\,\, \phi(\vect{n}),
\end{equation}
where $\vect{n}\equiv(n_1,n_2)=(b,a)/\sqrt{a^2+b^2}$ is the unit vector
orthogonal to $\vect{PQ}$ and
\begin{equation}
  \phi(\vect{n})= -n_1\ln \frac{n_1}{n_1+n_2} -n_2\ln \frac{n_2}{n_1+n_2}.
\end{equation}

Heuristically one expects that in the limit $E\rightarrow\infty,
h\rightarrow\infty,W_h\rightarrow\infty$, the profile of the Young diagram
will be described by a smooth curve $y=y(x)$ where $y=h/\sqrt{E}$ and
$x=W_h/\sqrt{E}$ are the scaling variables. The normal vector can be
parameterized as
\begin{equation*}
  \vect{n}=\left(-\frac{y'(x)}{\sqrt{1+y'^2(x)}},
    \frac{1}{\sqrt{1+y'^2(x)}} \right).
\end{equation*}
Therefore
\begin{equation}
  \phi(\vect{n})=\frac{y'(x)}{\sqrt{1+y'^2(x)}}
  \ln\left[-\frac{y'(x)}{1-y'(x)} \right]
-\frac{1}{\sqrt{1+y'^2(x)}}
  \ln\left[\frac{1}{1-y'(x)} \right].
\end{equation}

In the lattice model, the points $P$ and $Q$ were taken to be far apart.
However in the new scale $(x,y)$ one now assumes that they are close enough
in order to ensure that the interface is locally flat.  The total number of
Young diagrams $\Omega$ with a given area $E$ is obtained by adding all
such local configuration, i.e.
\begin{equation}
  \Omega =\exp\left(\sqrt{E} \int_0^\infty 
    \rmd x \sqrt{1+y'^2(x)}\, \phi(\vect{n})\right),
\label{omega}
\end{equation}
with the area constraint 
\begin{equation}
\int_0^\infty y(x) \rmd x =1.
\label{constraint}
\end{equation}

For large $E$, the most dominant contribution to $\Omega$ arises from the
optimal curve $y=y(x)$ which maximizes the integral in \eref{omega} with
the constraint \eref{constraint}. This optimal curve describe the limit
shape of the Young diagrams.  Thus we are led to the variational problem of
extremizing
\begin{equation}
  \mathcal{L}_0
  =\int_0^\infty \rmd x \left[ 
    y'(x)
    \ln \frac{-y'(x)}{1-y'(x)} 
    -
    \ln \frac{1}{1-y'(x)} 
  \right] -\lambda\int_0^\infty y(x)\, \rmd x,
\end{equation}
where $\lambda$ is a Lagrange multiplier.  This leads to the
Euler-Lagrange equation, which simplifies to
\begin{equation}
  \cfrac[l]{\rmd}{\rmd x} \ln \frac{-y'(x)}{1-y'(x)} =-\lambda. 
\end{equation}
We solve this equation with the boundary conditions $y(\infty)=0$ and
$y(x\rightarrow 0)\rightarrow\infty$.  The later condition follows from the
fact that $y\equiv h/\sqrt{E}\sim \ln E$ when $x\equiv
W_h/\sqrt{E}\rightarrow 0$ for large $E$~\cite{Erdos}. Therefore $y(0)$
diverges in the limit $E\rightarrow\infty$.  The solution gives the
equation of the limiting shape as
\begin{equation}
  y(x)=-\frac{1}{\lambda} \ln \bigl(1-\rme^{-\lambda x}\bigr)
\quad\mbox{with}\quad
\lambda=\frac{\pi}{\sqrt 6},
\end{equation}
where $\lambda$ is obtained by using the constraint \eref{constraint}.

The goal of this paper is to extend this derivation to the \MDP with $p>0$.
This is a priori non-trivial since now one has to take into account the
restriction on the steps. In the following we will use an exact
correspondence between \MDP with $p>0$ and a unrestricted partition
($p=0$).

\begin{figure}
\centering
\includegraphics[height=7cm]{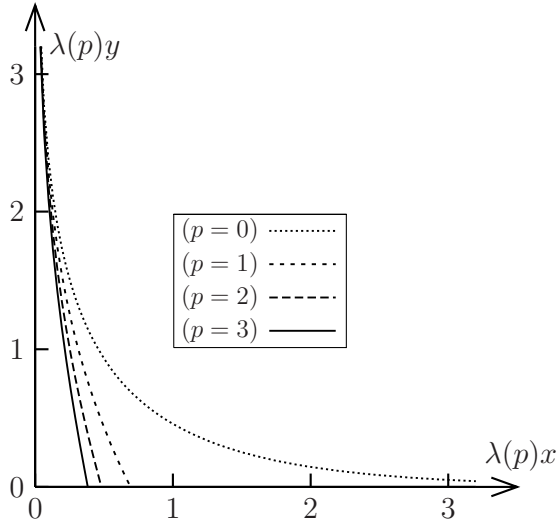}
\caption{\label{shapefig} Limit shapes for the minimal difference $p$
  partitions with $p=0,1,2$, and $3$, where $\lambda(0)=\pi/\sqrt{6}$,
  $\lambda(1)=\pi/\sqrt{12}$, $\lambda(2)=\pi/\sqrt{15}$, and
  $\lambda(3)=0.752617\ldots$.}
\end{figure}

Let $\{h_j\}$ denote the set of non-zero heights in a given unrestricted
partition ($p=0$) $E=\sum_{j=1}^N h_j$, where $h_j\ge h_{j+1}$ for all
$j=1,2,\ldots,N-1$. Let us now define a new set of heights
$h'_j=h_j+p(N-j)$ for $j=1,2,\ldots,N$. Thus $h'_j-h'_{j+1} =
h_j-h_{j+1}+p$ for all $j=1,2,\ldots,N-1$ and $h'_N=h_N>0$. Since
$h_j-h_{j+1} \ge 0$, the new heights thus satisfy the constraint
$h'_j-h'_{j+1} \ge p$ for all $j=1,2,\ldots,N-1$. Since the mapping is one
to one, the total number of local \MDP configuration satisfies
\begin{equation*}
\Omega_p(i,h'_i;j,h'_j)=\Omega_0(i,h_i;j,h_j).
\end{equation*}
Moreover, $h_i-h_j=h'_i-h'_j-p(j-i)$. Therefore using \eref{omega0},
\begin{equation*}
  \Omega_p=\binom{(j-i)(1-p)+h'_i-h'_j}{j-i}.
\end{equation*}
The fact that the mapping does not preserve the total area does not spoil
the argument since here we only deal with local \MDP configurations. The
area constraint is a global one which is implemented at the end of the
calculation via a Lagrange multiplier. Following the same steps as before
we arrive at the variational problem of extremizing
\begin{equation}
  \mathcal{L}_p =\int_0^\infty \rmd x \left[ \bigl(p+y'(x)\bigr) \ln
    \frac{-p-y'(x)}{1-p-y'(x)} - \ln \frac{1}{1-p-y'(x)}
  \right]
  -\lambda\int_0^\infty y(x)\, \rmd x.
\end{equation}
Using the same Euler-Lagrange formalism, finally leads us to the equation
of the limit shape for $p>0$,
\begin{equation}
  y=-\frac{1}{\lambda} \ln (1-\rme^{-\lambda x}) -px.
\label{limit}
\end{equation}
The Lagrange multiplier $\lambda$ in \eref{limit} can be determined by
using condition $y\ge 0$ and the normalization $\int_0^{ x_m} y(x)\, \rmd x
=1$, where $ x_m$ is the solution of the equation $y( x_m)=0$.  Writing
$\exp (x_m) = y^*$, it satisfies $y^* -y^{* 1-p}=1$, and in terms of $y^*$
one finds
\begin{equation}
  \lambda^2\equiv \lambda^2(p)= \frac{\pi^2}{6} -\dilog (1/y^*)-
  \frac{p}{2} (\ln y^*)^2,
\label{b(p)}
\end{equation}
where $\dilog(z)=\sum_{k=1}^\infty z^k k^{-2}$ is the dilogarithm function.
$\lambda(p)$ is a constant which depends on the parameter $p$.  For
instance for $p=0,1$ and $2$, one finds $\lambda(0)=\pi/\sqrt{6}$,
$\lambda(1)=\pi/\sqrt{12}$ and $\lambda(2)=\pi/\sqrt{15}$ in agreement with
the earlier known results \cite {Vershik, Romik}.  \Fref{shapefig} shows
the limit shapes for the \MDP with $p=0,1,2$, and $3$.

\Eref{limit} implies that the inverse function $x(y)=\lambda^{-1} \ln
\phi(y)$ satisfies 
\begin{equation}
  \phi(y)-\rme^{-\lambda y} \phi(y)^{1-p}=1
\label{func}
\end{equation}
Amazingly this  equation appears in several apparently unrelated
contexts.

1. The generating function $S(t)=1+\sum_{k=1}^{\infty} s_k(q)t^k$ for the
number of connected clusters $s_k(q)$ of size $k$ in a q-ary tree 
satisfies~\cite{FE}
\begin{equation}
  S(t)-tS^{q}(t)=1.
\label{gene}
\end{equation}
This establishes a formal link between two different combinatorial objects,
on one hand the q-ary trees and on the other hand the \MDP problem with
$p=1-q$.  In graph theory~\cite {HP}, $s_k(q)$ is known as the generalized
Catalan number, which is given by 
\begin{equation*}
s_k (q)=\frac{1}{k} \binom{qk}{k-1}.
\end{equation*}

2. Consider the generating function of \MDP problem
$$Z(x,z)=\sum_E\sum_N \rho_p (E,N)x^{E}z^{N}$$ where $\rho_p(E,N)$ is the
total number of \MDP of $E$ in $N$ parts. In the limit $E\rightarrow
\infty$ the number of such partitions will be controlled by the
singularities of $Z(x,z)$ near $x=1$.  By setting $x=e^{-\beta}$, one gets
for $\beta \rightarrow 0 $ \cite {CMO}
\begin{equation}
  \ln Z(x,z)\rightarrow \int_0^{\infty} 
  \ln y_p\left(z\rme^{-\beta\epsilon}\right)\,
  \rmd\epsilon,
\label{Z}
\end{equation}
where the function $y_p(t)$  is
given by the solution of the equation
\begin{equation}
y_p(t)- t\, y_p^{1-p}(t)=1.
\label{fun1}
\end{equation}

3. In the physics literature \eref{func} also arises in the context
of exclusion statistics.  Exclusion statistics
\cite{Haldane,LLLanyons,Isakov,Wu,MS}---a generalization of Bose and Fermi
statistics---can be defined in the following thermodynamical sense. Let
$Z(\beta, z)$ denote the grand partition function of a quantum gas of
particles at inverse temperature $\beta$ and fugacity $z$. Such a gas is
said to obey exclusion statistics with parameter $0\le p\le 1$, if
$Z(\beta, z)$ can be expressed as an integral representation
\begin{equation}
  \ln Z(\beta,z)=\int_0^{\infty} {\tilde \rho}(\epsilon)
  \ln y_p\left(z\rme^{-\beta\epsilon}\right)\,
  \rmd\epsilon,
\label{thermo}
\end{equation}
where ${\tilde \rho}(\epsilon)$ denotes a single particle density of states
and the function $y_p(t)$, which encodes fractional statistics is given by
the solution of \eref{func}. Well known microscopic
quantum mechanical realizations of exclusion statistics are the Lowest
Landau Level (LLL) anyon model \cite{LLLanyons} and the Calogero model
\cite{Isakov}, with ${\tilde \rho}(\epsilon)$ being, respectively, the LLL
density of states and the free one dimensional density of states.

The fact that the same  equation appears in all three cases is
obviously not fortuitous.  The link between 2 and 3 follows from the fact
that exclusion statistics has a combinatorial interpretation in terms of
minimal difference partitions which generalizes the usual combinatorial
interpretation of Bose statistics (resp Fermi) in terms of partitions
without (with) restrictions. Let us briefly recall this correspondence. Let
$n_i$ be the number of columns of height $h=i$ in a Young diagram of a
given partition of $E$, then $E=\sum_i n_i\epsilon_i$ can be interpreted as
the total energy of a non-interacting quantum gas of bosons where
$\epsilon_i=i$ for $i=1,2,\ldots,\infty$ represent equidistant single
particle energy levels and $n_i=0,1,2,\ldots,\infty$ represents the
occupation number of the $i$-th level.  If one now puts the restriction
that $h_j > h_{j+1}$ (e.g. allowed partitions of 4 are: $4$ and $3+1$),
then the restricted partition problem corresponds to a non-interacting
quantum gas of fermions, for which $n_i=0,1$. If in addition, one restricts
the number of summands to be $N$, then clearly $N=\sum_i n_i$ represents
the total number of particles. For example, if $E=4$ and $N=2$, the allowed
partitions are $3+1$ and $2+2$ in the unrestricted problem, whereas the
only allowed restricted partition is $3+1$.  The number $\rho(E,N)$ of ways
of partitioning $E$ into $N$ parts is simply the micro-canonical partition
function of a gas of quantum particles with total energy $E$ and total
number of particles $N$:
\begin{equation}
  \rho(E,N) =\sum_{\{n_i\}} \delta\left(E-\sum_{i=1}^{\infty} n_i
    \epsilon_i\right)\,
  \delta\left(N-\sum_{i=1}^{\infty} n_i\right).
\label{ren1}
\end{equation}
For both unrestricted and restricted partitions, one can readily check that
the grand partition function $Z(e^{-\beta},z)=\sum_N\sum_E z^N \rme^{-\beta
  E}\rho(E,N)$, in the limit $\beta\rightarrow 0$, is nothing but the one
in \eref{Z}, with $p=0$ and $p=1$ respectively.

For a quantum gas obeying exclusion statistics with parameter $p$ it is a
priori not obvious how to provide a combinatorial interpretation since the
underlying physical models with exclusion statistics describe interacting
models. However in some specific cases , such as the Calogero model one can
show that the spectrum can be parameterized as a free spectrum with some
restrictions on the quantum numbers which reflect the fact that the Pauli
principle is replaced by a stronger exclusion principle \cite{VP, AP}. This
exclusion is enforced at the level of the Young diagrams by the constraint
$h_{j} -h_{j+1} \ge p$. The link between 1 and 3 expresses this
correspondence in terms of counting of states. Exclusion statistics can be
implemented by putting $n$ particles in $m$ sites on a one-dimensional
lattice, under the restriction that any two particles be at least $p$ sites
apart. For a periodic lattice, the number of ways of doing the above
is~\cite{AP2}
\begin{equation*}
D_{m,n}=\frac{m\Gamma\bigl(m+(1-p)n\bigr)}{\Gamma\bigl(n+1\bigr)
  \Gamma\bigl(m+1-pn\bigr)}.
\end{equation*}
One can check that $D_{1,n}= s_n(1-p)$ which
allows to interpret the generalized Catalan numbers as quantum degeneracy
factors.

We acknowledge the support of the Indo-French Centre for the Promotion of
Advanced Research (IFCPAR/CEFIPRA) under Project 3404-2.

\end{document}